# Visualizationof Job Scheduling in Grid Computers

M. A. Awad
Faculty of Computer & Information
Mansoura University, Egypt
m.3abdelhady@gmail.com

M. Z. Rashad
Faculty of Computer & Information
Mansoura University, Egypt
magdi_z2011@yahoo.com

M. A. Elsoud
Faculty of Computer & Information
Mansoura University, Egypt
moh_soud@mans.edu.eg

M. A. El-dosuky
Faculty of Computer & Information
Mansoura University, Egypt
mouh_sal_010@mans.edu.eg

**ABSTRACT**
One of the hot problems in grid computing is job scheduling. It is known that the job scheduling is NP-complete, and thus the use of heuristics is the de facto approach to deal with this practice in its difficulty. The proposed is an imagination to fish swarm, job dispatcher and Visualization gridsim to execute some jobs.

## 1. INTRODUCTION
Grid computing[1][2] is a type of distributed computing that work to divide computing, applications, and data storage, or network resources between organizations and geographical spread, And that its road to change the method of dealing with complex computational problems. In a large scale computing systems such as grid computing systems, there are often large amounts of resources available to be used for computing jobs. Since these resources can cost many millions of dollars to achieve the maximum made? use of an important problem. Scheduling in a grid computing [3] [4] system is not as simple as scheduling on a multiprocessor system because of several factors. These factors include the fact that grid resources are sometimes used by paying customers who have interest in how their jobs are being scheduled. Also, grid computing systems usually operate in very remote locations Task Scheduler [5] for groups can occur across a network. Because of these reasons considering scheduling in grid computing is interesting and important problem to examine.

## 2. ARTIFICIALFISH SWARM ALGORITHM
The basic idea of AFSA [6] [7] is simulated fish behaviors such as swarming, preying and following with local search of fish individual for reaching the global optimum; it is random and parallel search algorithm.
The Artificial Fish [8] [9] [10] (AFunderstands the external perception of vision. *X*is the current state of an AF, *Visual* is the visual distance, and *Xv*is the visual position at some moment. If the state at the visual position is better than the current state, it goes forward a step in this direction, and arrives the *Xnext* state; otherwise, continues an inspecting tour in the vision. The greater number of inspecting tour the AF does, the more knowledge about overall states of the vision the AF obtains. Certainly, it does not need to travel throughout complex or infinite states, which is helpful to find the global optimum by allowing certain local optimum with some uncertainty.

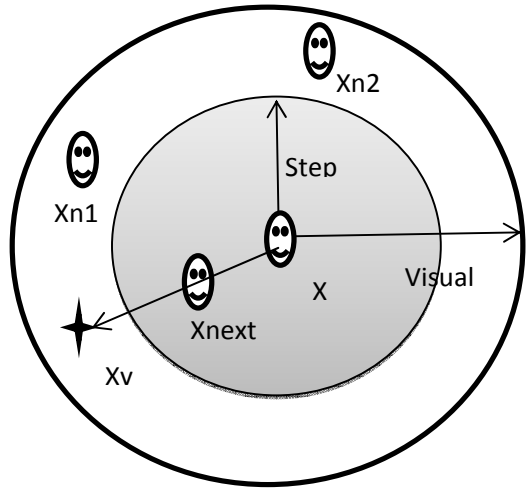

**Fig.1 Vision concept of Artificial Fish**

Let X= $(x_1, x_2, \ldots, x_n)$ and Xv=$(x_{1v}, x_{2v}, \ldots, x_{nv})$, then process can be expressed as follows:

$$x_i^v = x_i + Visual.Rand(), i\ (0, n] \quad (1)$$

$$X_{next} = X + \frac{X_v - X}{||X_v - X||} Step.Rand() \quad (2)$$

Where Rand () produces random numbers between 0 and 1, Step is the step length, and xi is the optimizing variable, n is the number of variables. The AF model includes two parts (variables and functions). The variables include: X is the current position of the AF, Step is the moving step length, Visual represents the visual distance, try_number is the try number and $\delta$s the crowd factor (0 <$\delta$< 1). The functions include the behaviors of the AF: AF_Prey, AF_Swarm, AF_Follow, AF_Move.

## 3. PREVIOUS WORK

### 3.1. *Gridsim*[11]:
is a platform for solving large problems in science and engineering. The scheduling and management of resourcesin such a distributed systems are complex and for that, they require advanced demands tools for analysing the algorithms before implementing them to the real systems. Simulation seems to be the only feasible way to analyze algorithms on distribution systems of heterogeneous resources. Simulations are also effective in working with very large virtual problems that would require the implicated of a big number of users and resources.

### 3.2. SimGrid[12]:
A Toolkit provides the basic functions for simulation of distributed applications in heterogeneous distributed environments. The specific objective of the project is to facilitate research in the field of distributed and parallel scheduling application on distributed computing platforms from simple network of workstations for computer networks.

### 3.3. *GridLAB-D*[13]:
Is a flexible simulation environment that can be integrated with a variety of third-party data management and analysis tools. The core of GridLAB-D has an advanced algorithm that simultaneously coordinates the state of millions of independent devices, each of which is described by multiple differential equations.

### 3.4. *Alchemi*[14]:
The framework is designedin order to make the grid construction and grid software development as easy as possible without sacrificing flexibility, scalability and reliability, allowing you to assemble the power of networked machines into a virtual supercomputer (desktop grid).

### 3.5. *GAT*:
The Grid Application Toolkit GAT provides unified and simple programming interface for Gridinfrastructure, tailored to the needs of Grid application programmers and users. There will be a need for unified programming interface for application developers to create a new generation of applications. Each of implementation of GAT deals with the complexity and variety of middleware services based network via the so-called adapters.

### 3.6. *Globus*[15]:
Toolkit is designed to enable people to create computer networks. Globus is an open-source initiative aimed at creating new networks capable of computing scale seen only in supercomputers yet.

## 4. PROPOSED FRAMERORK

### *4.1. Fish swarm*

A lot has been heard about fish swarm algorithm, but so far, no application shows how the fish moves. The proposed application is to visualize a number of fish that swarm in random way and can add fish and each fish carries a task to execute.

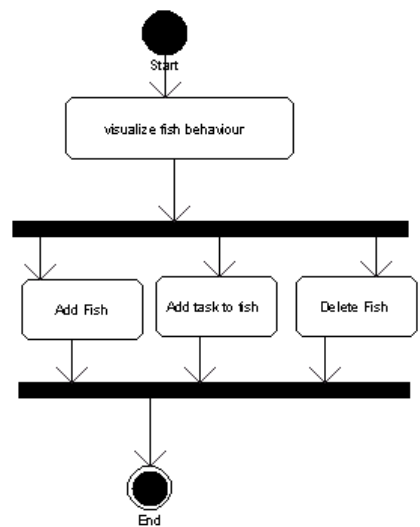

**Fig 2:Fish swarm simulatorDiagram**

Fish number can be specified, also fish task and ability to delete fish or add new fish with a new task

### *4.2. Dispatcher*

The dispatcher will work with other members of the distribution team to increase company profitability and customer satisfaction by efficient and cost-effective scheduling of deliveries and routing of trucks.

The dispatcher determines the coordinates of each task based on keywords and this will reduce dimensionality which saves the time for clustering and classification tasks.

Input: There will be a folder known as Fields, and inside the folder for each field (Math, texts, etc.)and within each field of text files (notepad). The name of text file is a keywords or operations and there is another folder named Item, inside the item folder there are text files, each file is a task, and contains a set of keywords names and the domain name which specializes the task.

Output: Software is required to calculate each item Coordinates (x, y) based on the names of the keywords

Example: Assuming that keywords in the field are:

| j | i | h | j | f | e | d | c | b | a |

Assuming that keywords in the message are:

| h | f | c | b | A |
|---|---|---|---|---|

A value of 1 to express a word and a value of 0 for their absence:

| j | i | h | j | f | e | d | c | b | a |
|---|---|---|---|---|---|---|---|---|---|
| 0 | 0 | 1 | 0 | 1 | 0 | 0 | 1 | 1 | 1 |

Because the number of words will be very large, a division for the previous matrix into two or four matrixes is proposed as the following:

First part (lower):

| e | d | c | b | a |
|---|---|---|---|---|
| 0 | 0 | 1 | 1 | 1 |

1+2+4 = 7

Second part (upper):

| j | i | h | j | f |
|---|---|---|---|---|
| 0 | 0 | 1 | 0 | 1 |

1+4=5

Calculated values will be transferred from binary to decimal system for each matrix.

(00111) binary= (7) decimal

The remaining two digits could be used as the circle center coordinates that represents the task(x,y).

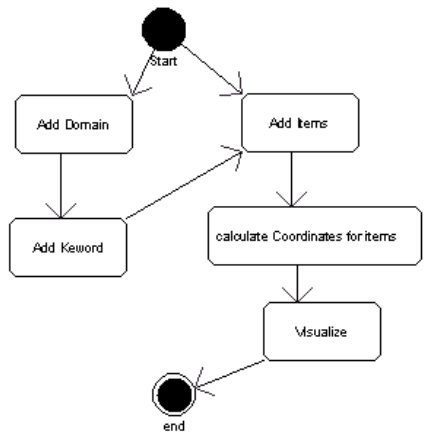

**Fig 3: Dispatcher Diagram**

## 4.3. Grid Simulator

The Grid Simulator toolkit allows modeling and simulating of entities in parallel and distributed computing (PDC) systems-users, applications, and resource broker (schedulers) for design and evaluation of scheduling algorithms. A resource can be a single processor or multi-processor with shared or distributed memory and managed by time or space shared schedulers. The processing nodes within a resource can be heterogeneous in terms of processing capability, configuration, and availability.

The idea of this software is to visualize simulate grid and make easy to add resources and number of jobs and show the Statistics about each job and resources. First: the number of users and the number of resources to work with should be specified. Second: number of machines should be specified for each resource and number of P.E for each machine, also P.E rating and number of jobs for each user.

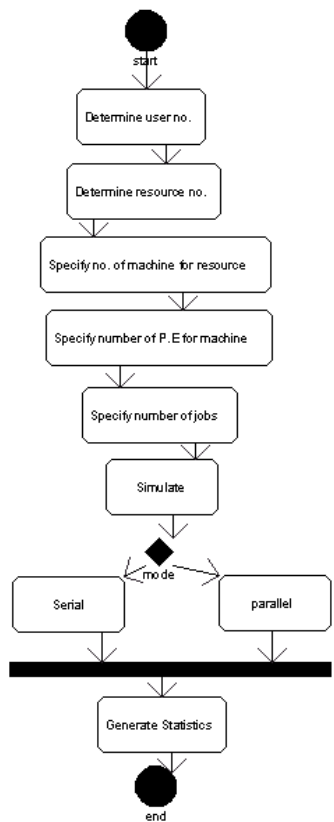

**Fig 4: Grid Simulator**

**Table 1. Comparison between simgrid, gridsim and proposed system**

|  | **SimGrid** | **GridSim** | **Proposed system** |
|---|---|---|---|
| CPU | coarse d.e.s. | coarse d.e.s. | coarse d.e.s. |
| Disk | - | fine d.e.s. | fine d.e.s. |
| Network | math/d.e.s. | fine d.e.s. | fine d.e.s. |
| Application | d.e.s./emul | coarse d.e.s. | coarse d.e.s. |
| Requirement | C or JAVA | JAVA | JAVA |
| Setting | controlled | controlled | controlled |
| Scale | few 10 000 | few 100 | few 100 |
| GUI | - | - | yes |

## 5. CONCLUSION

This paper presents three frameworks, one for visualizing the behavior of fish swarm (The basic idea of AFSA is simulated fish behaviors such as swarming, preying, following with local search of fish individual for reaching the global optimum; it is random and parallel search algorithm). The second for visualizing grid simulator that uses gridsim framework as grid simulator (the ability to add resource, user and number of tasks has been added, and represent information about each task such as execution time, waiting time ,.. Etc.), also comparison between this and girdsim and grid simulator. The third framework is dispatcher that determine the coordinates of each task Based on keywords and that will reduce dimensionality which saves time for clustering and classification tasks.

## 6. FUTURE WORK

Proposed framework that explained above will developed in a cloud to be accessible for all